\documentclass[pra,showpacs,twocolumn,10pt,aps,superscriptaddress]{revtex4-1}
\usepackage[T1]{fontenc}
\usepackage[utf8]{inputenc}
\usepackage[english]{babel}
\usepackage{amsmath}
\usepackage{amssymb}
\usepackage{amsthm}
\usepackage{dsfont}
\usepackage{amsfonts}
\usepackage{xcolor}
\usepackage{graphicx}
\usepackage{array}
\usepackage{ragged2e}
\usepackage{booktabs}
\usepackage{url}
\usepackage{dsfont}
\usepackage{tcolorbox}
\usepackage{url}
\usepackage[linktocpage, colorlinks=true, linkcolor=darkblue, citecolor=darkblue, urlcolor=darkblue]{hyperref}
 \definecolor{darkblue}{rgb}{0,0,.5}
\usepackage{dcolumn}

\newcommand{\bra}[1]{\langle #1|}
\newcommand{\ket}[1]{|#1\rangle}
\newcommand{\braket}[2]{\langle #1|#2 \rangle}
\newcommand{\av}[1]{\langle #1 \rangle}
\newcommand{\avin}[3]{\langle #1|#2|#3 \rangle}

\newcommand{\up}{\uparrow}
\newcommand{\down}{\downarrow}

\makeatletter
\newcommand*{\balancecolsandclearpage}{%
	\close@column@grid
	\clearpage
	\twocolumngrid
}
\makeatother

\begin{document}
	\title{Polaritonic Coupled-Cluster Theory}
	\author{Uliana Mordovina}
	\email{uliana.mordovina@mpsd.mpg.de}
	\affiliation{Max Planck Institute for the Structure and Dynamics of Matter, Center for Free Electron Laser Science, Luruper Chaussee 149, 22761 Hamburg, Germany}
	\author{Callum Bungey}
	\affiliation{Centre for Computational Chemistry, School of Chemistry, University of Bristol, Bristol BS8 1TS, United Kingdom}
	\author{Heiko Appel}
	\affiliation{Max Planck Institute for the Structure and Dynamics of Matter, Center for Free Electron Laser Science, Luruper Chaussee 149, 22761 Hamburg, Germany}
	\author{Peter J. Knowles}
	\affiliation{School of Chemistry, Cardiff University, Main Building, Park Place, Cardiff CF10 3AT, United Kingdom}
	\author{Angel Rubio}
	\affiliation{Max Planck Institute for the Structure and Dynamics of Matter, Center for Free Electron Laser Science, Luruper Chaussee 149, 22761 Hamburg, Germany}
	\affiliation{Center for Computational Quantum Physics (CCQ), Flatiron Institute, 162 Fifth Avenue, New York NY 10010}
	\author{Frederick R. Manby}
	\email{fred.manby@bristol.ac.uk}
	\affiliation{Centre for Computational Chemistry, School of Chemistry, University of Bristol, Bristol BS8 1TS, United Kingdom}
	\date{\today}
	
	\begin{abstract}
		We develop coupled-cluster theory for systems of electrons strongly coupled to photons, providing a promising theoretical tool in polaritonic chemistry with a perspective of application to all types of fermion-boson coupled systems. We show benchmark results for model molecular Hamiltonians coupled to cavity photons. By comparing to full configuration interaction results for various ground-state properties and optical spectra, we demonstrate that our method captures all key features present in the exact reference, including Rabi splittings and multi-photon processes. Further, a path on how to incorporate our bosonic extension of coupled-cluster theory into existing quantum chemistry programs is given. 
	\end{abstract}
	
	\maketitle

	In recent years seminal experiments at the interface between quantum optics, quantum chemistry and material sciences have shown that when photons and matter couple strongly the emergence of light-matter hybrid states, called polaritons, can substantially change chemical and physical properties of molecular systems \cite{kena2010room,hutchison2012modifying,coles2014strong,orgiu2015conductivity,chikkaraddy2016single_strong,ebbesen2016hybrid,sukharev2017optics,zhong2017long-range2,PhysRevLett.122.173902}. This has led to the observation of changes in chemical reactions \cite{thomas2016ground,lather2019cavity}, suppression of photon-degradation and photobleaching \cite{peters2019effect,munkhbat2018suppression}, tunable third-harmonic generation from polaritons \cite{Barachati2018a}, room-temperature superfluidity in a polariton condensate \cite{Lerario2017}, or modifications of intersystem crossings \cite{stranius2018selective}. While experimentally the influence of strong coupling on matter, e.g., due to placing molecules inside a cavity or plasmonic nanostructure, has been firmly established, theoretical approaches to describe situations where photonic, electronic and nuclear degrees of freedom become strongly mixed
	so far do not provide a detailed and general explanation of the observed effects \cite{ruggiNatureSpec2018,kockum2019ultrastrong,flick2018strong,ribeiro2018polariton}.
	
	Although many observations can be described by problem-adopted quantum-optical models \cite{feist2015extraordinary,schachenmayer2015cavity,herrera2016cavity,*herrera2017dark,zeb2017exact,reitz2019langevin,strathearn2018efficient,f2018theory,RevModPhys.91.025005},
	\emph{ab initio} methods are necessary for a detailed and unbiased understanding of the effects \cite{flickPNAS2017,ruggiNatureSpec2018,schafer2019modification}. To this aim, some electronic structure methods have already been extended to include
	the photons explicitly \cite{galego2015cavity,kowalewski2016non,de2016unified,triana2019revealing,vendrell2018collective,flick2018cavity,rivera2019variational,Karlsson2018,delPino2018tensor, schafer2019modification} with quantum-elctrondynamical density functional theory (QEDFT) being one of the most prominent approaches \cite{ruggenthaler2014quantum,pellegrini2015optimized,flick2015kohn,flickPNAS2017,flick2018light}. While being formally exact, QEDFT relies on development of accurate and robust approximate functionals, which is especially challenging in case of significant correlation effects \cite{cohen2011challenges} and strong matter-cavity couplings \cite{flick2018ab-initio}.
	
	When reliability and accuracy is concerned, coupled-cluster (CC) theory \cite{vcivzek1971correlation, bartlett2007coupled, crawford2000introduction,riplinger2013efficient} has become the method of choice in quantum chemistry.
	This wave-function method provides a hierarchy of approximations with truncation based on excitation order from a mean-field reference state. Size-consistent and size-extensive molecular electronic energies, as well as other ground and excited-state properties, can be calculated with chemical accuracy \cite{lee1995achieving}. This makes an extension of CC theory to the case of strong matter-photon coupling highly desirable.
	Fortunately, there is nothing intrinsic in CC theory that imposes a restriction to purely electronic problems. The requirements for a computationally tractable CC theory can be satisfied for bosonic degrees of freedom as well, as demonstrated by pioneering applications to molecular vibrations \cite{ove2004vcc}.
	
	In this Letter, we develop the theoretical framework for CC calculations on systems of interacting electrons coupled to photons, 
	and illustrate the potential of such an approach to describe the behavior of matter strongly coupled to cavity modes with a computational cost that scales polynomially with problem size.
	
	We consider fixed-nucleus molecular Hamiltonians coupled to a cavity in the dipole limit \cite{craig1998molecular,spohn2004dynamics,rokaj2018light}
	\begin{align} \label{eq:Hgen}
	\hat{H}  
	&=  \hat{H}_\mathrm{e} + \sum\limits_{\alpha} \omega_{\mathrm c,\alpha} \hat{a}^\dagger_\alpha \hat{a}_\alpha + \gamma_\alpha\omega_{\mathrm c,\alpha} \hat{d} \left( \hat{a}^\dagger_\alpha + \hat{a}_\alpha \right) + \gamma^2_\alpha\omega_{\mathrm c,\alpha} \hat{d}^2,
	\end{align}
	where $\hat H_\mathrm e$ is the electronic Hamiltonian, $\hat{a}^{(\dagger)}_\alpha$ are the creation and annihilation operators for a cavity mode with frequency $\omega_{\mathrm c,\,\alpha}$, and $\hat{d}$ is the electronic dipole operator.
	The coupling parameter $\gamma_\alpha$ tunes the strength of the light-matter interaction;
	here we focus mostly on cases where $\gamma_\alpha > 0.05$, where the system is typically considered to be in the strong-coupling regime \cite{flick2018strong, galego2015cavity}.
	The theory can be extended to more general light-matter Hamiltonians \cite{craig1998molecular,spohn2004dynamics,flickPNAS2017,ruggiNatureSpec2018} in a straight-forward manner.
	
	CC theory is based on an exponential ansatz for the ground-state wavefunction
	\begin{equation}\label{eq:CCansatz}
	\ket{\Psi^{\rm CC}_0} = e^{\hat{T}} \ket{\Phi_0},
	\end{equation}
	where $\ket{\Phi_{0}}$ is an uncorrelated reference state (usually Hartree-Fock) and $\hat{T}$ is the cluster operator. The cluster operator is a weighted sum of excitation operators
	\begin{equation}\label{eq:closterOp}
	\hat T = \sum_\mu t_\mu \hat\tau_\mu \;,
	\end{equation}
	where $\mu$ labels a general excitation in the system, and the $t_\mu$ amplitudes are to be determined. The exponential form makes the CC state multiplicatively separable, introducing the fundamental feature of size-extensivity. It also has the property that even when the cluster operator is truncated to include only low-order excitations, higher-order effects are incorporated through the expansion of the exponential of $\hat T$.
%
		
	Standard CC theories are classified by the number of electrons excited in the list of operators $\tau_\mu$. For example, the method that includes all single and double excitations is called CC singles and doubles (CCSD). The excitation operators are written in terms of fermionic creation and annihilation operators $\hat{c}^{(\dagger)}$, e.g.\ the operator $\hat\tau_i^a=\hat c_a^\dagger\hat c_i$ excites an electron from occupied orbital $i$ to unoccupied orbital $a$. 
	
	\begin{figure}[b]
		\includegraphics[width=.42\linewidth]{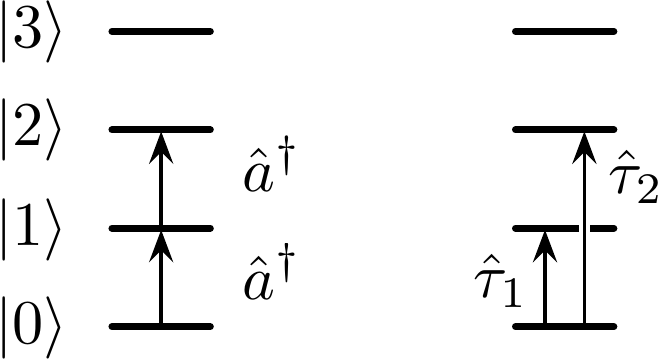} %
		\caption{Two types of excitation operators for a bosonic degree of freedom. Both can be used to access any number state, and both are sets of commutative operators. But the operators $\hat\tau_{n}$ are additionally nilpotent, simplifying CC formulations that use this form.}
		\label{fig:boson}
	\end{figure}
	
	The amplitudes and the ground-state energy are obtained through projected equations
	\begin{equation}
	\avin{\Phi_0}{ \hat{\bar{H}} }{\Phi_0} = E_0, \qquad
	\avin{\Phi_\mu}{ \hat{\bar{H}} }{\Phi_0} = 0, \label{eq:cc_eq}
	\end{equation}
	where $\ket{\Phi_\mu}=\hat\tau_\mu\ket{\Phi_0}$ and $\hat{\bar{H}}$ is the similarity-transformed Hamiltonian $e^{-\hat{T}} \hat{H} e^{\hat{T}}$. The polynomial scaling of the theory is attributed to the fact that usually a polynomial number of excitation operators enters $\hat{T}$ and that the Baker-Campbell-Hausdorff expansion of $\hat{\bar{H}}$ truncates \cite{crawford2000introduction}. The latter condition is met when excitation operators are commutative and nilpotent. 
	Hence, there is nothing special about electrons in the formulation and CC theory can be applied to general many-particle systems.
	
	The natural excitation operator for a bosonic mode is simply the creation operator $\hat a^\dagger$, but the lack of nilpotency ($(\hat a^\dagger)^2\ne0$) means that a formalism based on $\hat a^{\dagger}$ would lack the simple structure of electronic CC theory. However, if the number of photons in the system is limited to  $n_\mathrm{max}$, which is necessary for numerical treatment of bosons in the basis of Fock number states (see Supplemental Material for convergence tests), it is possible to map the bosonic mode to a lattice of $n_\mathrm{max}+1$ sites, $\ket 0,\ket 1,\dots,\ket{n_\mathrm{max}}$, each corresponding to a number state;
	then the excitation operators
	\begin{equation}\label{eq:rho_k}
	\hat{\tau}_n
	= \ket{n}\bra{0}
	\end{equation}
	clearly fulfill both the commutativity and nilpotency condition, while allowing any number state to be addressed (Fig.~\ref{fig:boson}).

	\begin{table*}[t]
		\caption{Ground-state energy $E_0$ and mode occupation $\av{\hat{a}^\dagger \hat{a}}$ of the four-site Hubbard chain coupled to one cavity mode (resonant to first bare absorption peak, see Appendix~\ref{app:conv_n}) with selected coupling strengths for different levels of CC theory compared with FCI as well as with FCI and CCSD for the bare electronic system (FCI(0) and CCSD(0)). We observe excellent agreement of CC results with FCI. 
			By including coupled excitations in the CC description the ground-state energy is improved. Further, purely photonic observables like mode occupation become accessible.
			Parameters: $\omega_c = 1.028$, $t_0 = 0.5$, $U= 1.0$, $d = [-1.5, -0.5, 0.5, 1.5]$, $n_\mathrm{max}=1$ (weak) $n_\mathrm{max}=4$ (strong) and $n_\mathrm{max}=7$ (ultra-strong).}
		\begin{ruledtabular}\label{tab:ground-state}
			\begin{tabular}{lcccccc}
				&\multicolumn{2}{c}{weak coupling}  &\multicolumn{2}{c}{strong coupling} &\multicolumn{2}{c}{ultra-strong coupling}\\
				&\multicolumn{2}{c}{$\gamma=0.01$ } &\multicolumn{2}{c}{$\gamma=0.07$ } &\multicolumn{2}{c}{$\gamma=0.2$ } \\
				\midrule
				& $E_0$ &  $\av{\hat{n}_{\rm p}}$ & $E_0$ &  $ \av{\hat{a}^\dagger \hat{a}}$ &  $E_0$ &   $\av{\hat{a}^\dagger \hat{a}}$\\
				\midrule 
				FCI(0)     & $-$1.43797 & ---                  &$-$1.43797 & ---                  &$-$1.43797 & --- \\
				CCSD(0)    & $-$1.43801 & ---                  &$-$1.43801 & ---                  &$-$1.43801 & --- \\
				CC-SD-S-0  & $-$1.43791 & 0                    &$-$1.43335 & 0                    &$-$1.40227 & 0 \\
				CC-SD-S-D  & $-$1.43795 & $2.14 \cdot 10^{-5}$ &$-$1.43551 & $1.04 \cdot 10^{-3}$ &$-$1.41745 & $7.75 \cdot 10^{-3}$ \\
				CC-SD-S-DT & $-$1.43796 & $2.24 \cdot 10^{-5}$ &$-$1.43561 & $1.09 \cdot 10^{-3}$ &$-$1.41873 & $8.57 \cdot 10^{-3}$ \\
				FCI        & $-$1.43792 & $2.27 \cdot 10^{-5}$ &$-$1.43557 & $1.11 \cdot 10^{-3}$ &$-$1.41864 & $8.69 \cdot 10^{-3}$
			\end{tabular}
		\end{ruledtabular}
	\end{table*}
	
	Now, to build CC theory for electron-photon systems, we introduce a more general cluster operator
	\begin{align}\label{eq:Tgen}
	\hat{T}& = 
	\sum_\mu t_\mu \hat{\tau}_\mu 
	+ \sum_{n} t_n \hat{\tau}_n
	+ \sum_{\tilde{\mu}, \tilde{n}} t_{\tilde{\mu} \tilde{n}} \hat{\tau}_{\tilde{\mu}} \hat{\tau}_{\tilde{n}}
	\end{align}
	in which the electronic terms $\hat\tau_\mu$ are supplemented by purely photonic excitations $\hat\tau_n$ and connected light-matter excitations $\hat{\tau}_{\tilde{\mu}} \hat{\tau}_{\tilde{n}}$. 
	
	\begin{figure}[t]
		\includegraphics[width=.6\linewidth]{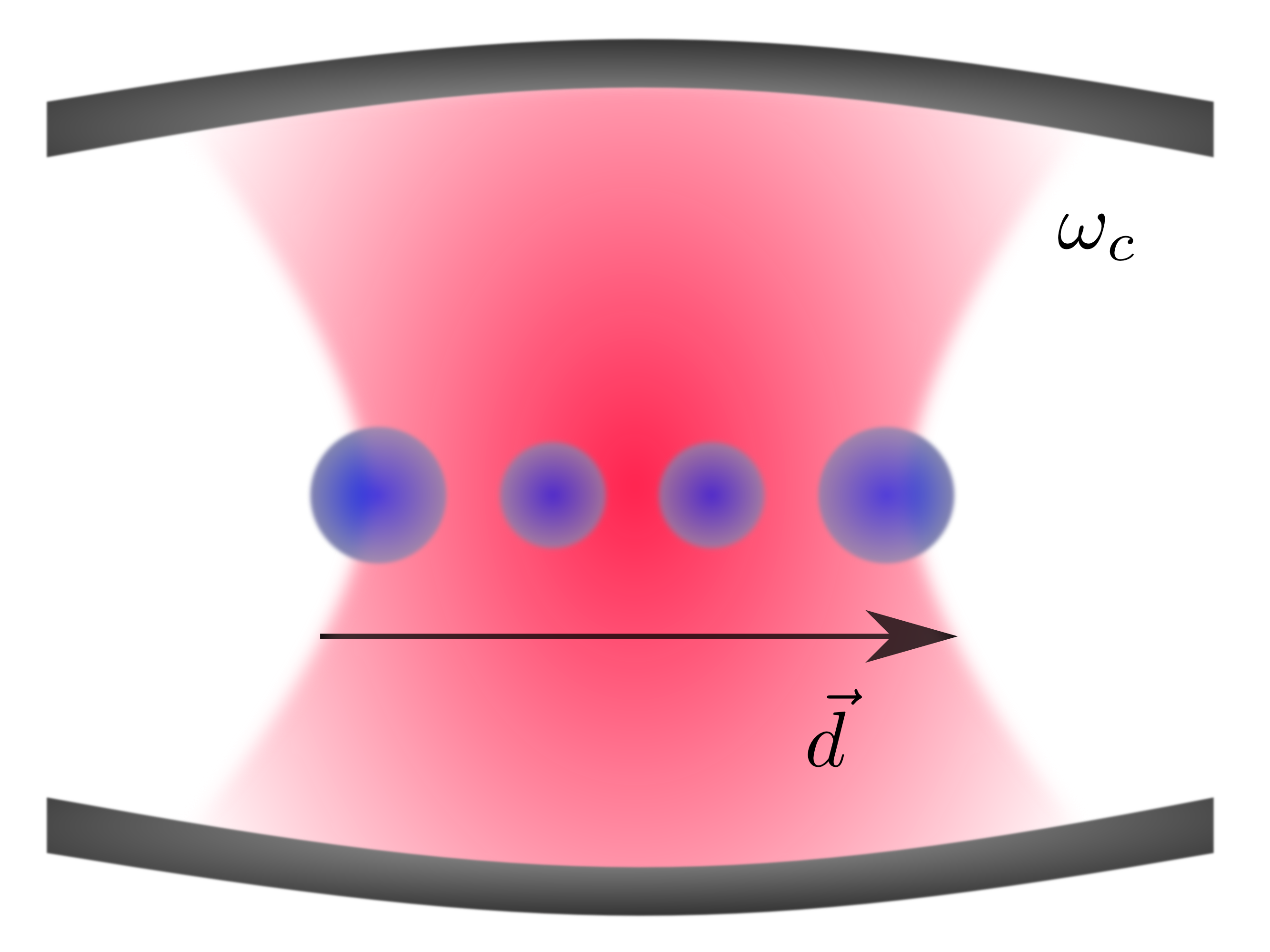} 
		\caption{A four-site Hubbard chain in half-filling serving as a model molecule. The molecule has a dipole moment $\vec{d}$, which is strongly coupled to the cavity mode with frequency $\omega_c$.}
		\label{fig:model}
	\end{figure}

	To describe the truncation of each term in the cluster operator, we extend the terminology common in electronic CC theory:
	in the acronym CC-$X$-$Y$-$Z$, $X$, $Y$, and $Z$ will specify the level of electronic, photonic, and mixed excitations respectively; CC-SD-S-0 refers to conventional electronic CCSD with additional single-photon excitations, CC-SD-S-D includes coupled excitations of one electron together with one photonic excitation additionally to CC-SD-S-0, and CC-SD-S-DT adds coupled double electronic together with one photonic excitations to CC-SD-S-D.
	In the one-mode calculations presented here the photonic excitation level is at most singles, but multi-photon excitations within that single mode are included (see Eq.~(\ref{eq:rho_k})). 
	
	As the reference state we take a product of an electronic Slater determinant and the vacuum state for radiation modes
	\begin{equation}\label{eq:el_ph_phi}
	\ket{\Phi_0} = \ket{\Phi} \otimes \ket{0} \;.
	\end{equation}
	In this form, polaritonic CC theory displays some similarities with the polaron ansatz in quantum optics \cite{PhysRevA.93.043843,PhysRevA.99.013807}. However, as opposed to CC theory, the polaron ansatz is developed for two-level systems and does not target the electronic structure explicitly.
	Additionally, since our method is in principle formulated in the full Hilbert space of the problem, it does not suffer per se from gauge-ambiguity issues exhibited by some effective quantum optical models \cite{PhysRevA.98.053819,di2019resolution}.
		
%
	
	Because the structure of polaritonic CC theory is closely analogous to that of conventional electronic CC theory, it is possible to anticipate many of the advantages of the formalism. The properties of the operators in Eq.~(\ref{eq:rho_k}) ensure truncation of the energy and amplitude equations to produce a polynomially scaling theory. Because the photon mode is modeled by a lattice, the situation is identical to one in which there is simply a single additional fermion beyond the $\uparrow$- and $\downarrow$-spin electrons, making an even more explicit connection to electronic CC theory. 	
	
	
	As a proof-of-principle, we consider a half-filled four-site Hubbard chain with an additional dipole coupled to a single photon cavity mode with frequency $\omega_c$ (see Fig.~\ref{fig:model} and Ref.~\cite{Dimitrov_2017}). Here we consider three values for the light-matter coupling parameter $\gamma$, representing weak ($\gamma=0.01$),  strong ($\gamma=0.07$), and ultra-strong ($\gamma=0.2$) coupling regimes \cite{flick2018strong,kockum2019ultrastrong}.
	The Hamiltonian of the model system is as in Eq.~(\ref{eq:Hgen}), but without the sum over $\alpha$ and with
	\begin{equation} 
	\hat H_{\rm e}  =  -t_0 \sum_{i\sigma} \left( \hat{c}^\dagger_{i+1,\sigma}\hat{c}_{i\sigma} + \hat{c}^\dagger_{i\sigma}\hat{c}_{i+1,\sigma} \right)
	+ U \sum_{i} \hat{n}_{i\up}\hat{n}_{i\down},
	\end{equation}
	where $\hat{n}_{i\sigma}=\hat{c}^\dagger_{i\sigma}\hat{c}_{i\sigma}$ denotes the density of a spin-$\sigma$ electron on site $i$, and $t_0$ and $U$ are the usual hopping and on-site repulsion constants.
	The dipole operator of the system is given by $\hat{d}=\sum_{i} d_i (\hat{n}_{i\up} + \hat{n}_{i\down})$.

	The results for the ground-state energy $E_0$ and photonic mode occupation $\av{\hat{a}^\dagger \hat{a}}$ in the system are summarized in
	Table \ref{tab:ground-state} for different levels of CC theory and compared to full configuration interaction (FCI) results for the coupled electron photon system and to FCI and CCSD results in the limit of vanishing electron-photon coupling, FCI(0) and CCSD(0), respectively.
	We observe an increasing impact of the cavity on the ground-state properties of the system for
	growing coupling strengths. For instance, the gap between CC-SD-S-0 and CCSD(0) energies
	widens due to increasing importance of the dipole self-interaction term. This behavior is captured very well with polaritonic CC theory for all coupling strength as soon as coupled excitation are included,
	namely with CC-SD-S-D and CC-SD-S-DT approximations.
	The photon-mode occupation $\av{\hat{a}^\dagger \hat{a}}$, which is now accessible with CC theory, is zero for CC-SD-S-0 for all coupling strength, which shows the intrinsic mean-field character of the CC-SD-S-0 approximation. The photon-mode occupation is also captured well as soon as coupled excitation are included. The best agreement with FCI is achieved for both observables with CC-SD-S-DT.
	
	However, a considerable effect of the cavity on the molecular ground state is observed only in the ultra-strong coupling regime ($\Delta E_0 \approx 0.02$), which is captured well with CC theories that include coupled excitations.
	For the other two regimes this impact is, as was to be expected, rather small \cite{flick2018ab-initio,schafer2018ab,schafer2019modification}.
	The power of CC theory therefore also lies in the treatment of excited (polaritonic) states that we show in the following.
	
	A key experimentally accessible feature to study is the ground-state absorption spectrum. The matter absorption cross section is given by \cite{rescigno1975rigorous,ruggiNatureSpec2018}
	\begin{equation}
	\label{eq:I}
	\sigma(\omega) =  4\pi \cfrac{\omega}{c} \, \mathrm{Im} \left[ \sum_k \cfrac{\left|\avin{\Psi_k}{\hat{d}}{\Psi_0}\right|^2}{(\omega_k - \omega_0) - \omega -  i\eta} \right],
	\end{equation}
	where $\ket{\Psi_k}$ are many-body eigenstates of $\hat{H}$ with energy $\hbar \omega_k$, $\omega$ is the frequency of incident light, and $\eta$ is a (small) broadening
	parameter accounting for the finite lifetime of the state.
	
		\begin{figure*}[t]
			\includegraphics[width=\textwidth]{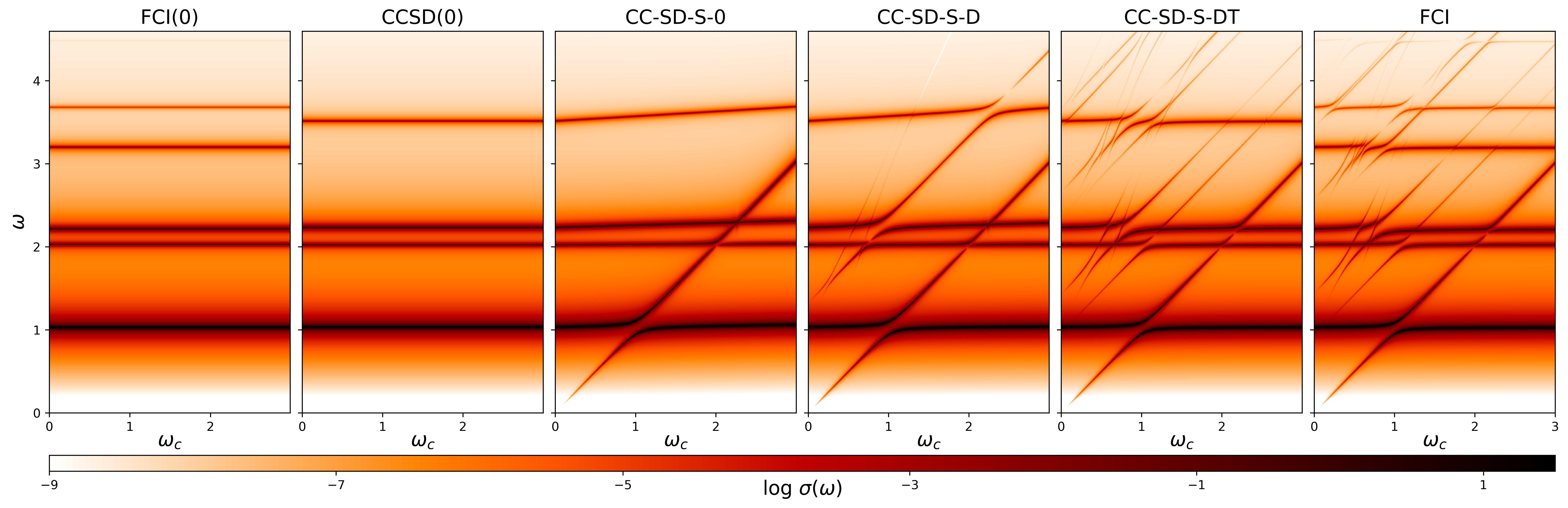}
			\caption{Ground-state absorption cross section $\sigma (\omega)$  of the half-filled four-site Hubbard chain in a cavity as a function of cavity frequency $\omega_c$ in the strong-coupling regime $\gamma=0.07$
			for different levels of CC theory compared with FCI results (right-hand side) and zero-coupling limit (FCI-0, left-hand side).
			Parameters: $t_0 = 0.5$, $U= 1.0$, $d = [-1.5, -0.5, 0.5, 1.5]$, $n_\mathrm{max}=4$, $\eta = 0.005$. \label{fig:spectra0-05} }
		\end{figure*}
	
	Here equation of motion CC (EOM-CC) theory will be used to access excited-state information \cite{geertsen1989equation,Stanton1993eom-cc,koch1994calculation,krylov2008equation}.
	In EOM-CC each excited state of the system is produced by applying a linear excitation operator to the correlated ground state:
	\begin{equation}
	\ket{\Psi^{\rm CC}_k} = \hat{\mathcal{R}}_k \ket{\Psi^{\rm CC}_0} 
	= e^{\hat{T}} \, \hat{\mathcal{R}}_k \ket{\Phi_0}.
	\label{eq:right_psi}
	\end{equation}
	The coefficients entering the excitation operator $\mathcal{R}_k$ are found by diagonalizing $\hat{\bar{H}}$ in the subspace of excited states $\ket{\Phi_\mu}$ addressed by each excitation operator in the expansion of $\hat{T}$ in Eq.(\ref{eq:Tgen}).
	Since $\hat{\bar{H}}$ is nonhermitian, we obtain a biorthogonal set of left ($\hat{\mathcal{L}}_j$) and right ($\hat{\mathcal{R}}_k$) operators satisfying
	\begin{align}
	\hat{\bar{H}}\hat{\mathcal{R}}_k & = E_k \hat{\mathcal{R}}_k, &
	\hat{\mathcal{L}}_j\hat{\bar{H}} & = \hat{\mathcal{L}}_j E_j, &
	\hat{\mathcal{L}}_j\hat{\mathcal{R}}_k & = \delta_{jk} \hat{\mathds{1}} ,
	\end{align}
	and we define the left eigenstates as 
	\begin{equation}
	\bra{\widetilde{\Psi}^{\rm CC}_j} = \bra{\Phi_0} e^{-\hat{T}} \hat{\mathcal{L}}_j.
	\end{equation}

	With these states, the absorption cross section can be approximated as
	\begin{equation}
	\label{eq:I_CC} 
	\sigma_\mathrm{CC} (\omega) = 4\pi \cfrac{\omega}{c} \,
	\mathrm{Im} \left[ \sum_k \cfrac{\avin{\widetilde{\Psi}^{\rm CC}_k}{ \hat{d} }{\Psi^{\rm CC}_0}\avin{\widetilde{\Psi}^{\rm CC}_0}{ \hat{d} }{\Psi^{\rm CC}_k}}{(\omega_k - \omega_0) - \omega -  i\eta} \right].
	\end{equation}

	A final detail is that the nonhermitian form of the CC theory can lead to problems for close-lying or degenerate states (such as at conical intersections), that either have complex eigenvalues or exhibit significant overlaps \emph{among} the right (or left) eigenvectors which can cause numerical instability. Both issues can be resolved following a correction method based on Ref.~\cite{kohn2007can} and details can be found in Appendix \ref{app:corr}.

	
	Leaving the weak-coupling case to Appendix \ref{app:weak}, we start our discussion of excited-state properties with the case of strong light-matter coupling of $\gamma=0.05$. Absorption spectra for different approximations are shown in Fig.~\ref{fig:spectra0-05} as functions of cavity frequency $\omega_c$.
	For reference, the panels at either end show the FCI spectrum for the purely electronic problem (left) and for the full light-matter problem (right). In between we show the results of EOM-CC calculations with various levels of truncation of the cluster operator.
	The bare electronic spectra (FCI(0) and CCSD(0)) feature horizontal non-dispersive absorption lines; as expected, approximate EOM-CC accurately reproduces low-energy features in the spectrum.
	
	When coupled to a cavity, the spectrum includes matter absorption lines and additional linear dispersive branches for one-photon (lines where $\omega=\omega_\mathrm c$) and many-photon processes ($\omega=2\omega_\mathrm c$, etc).
	In the strong-coupling case of Fig.~\ref{fig:spectra0-05},
	we see at most three-photon processes with significant amplitude. At points where there are matter excitations resonant with cavity modes we observe avoided crossings in the absorption lines.
	In these regions hybrid light-matter states (polaritons) are formed, accompanied by the signature Rabi splitting of an absorption peak. They can be clearly seen in various regions of the FCI spectrum, and are well captured by the EOM-CC approximation that includes coupled excitations. 
	The systematic improvement in the CC treatment as the cluster operator is extended is clear, and remaining deviations from FCI are largely caused by the truncation of the electronic
	part of the cluster operator.
	
	\begin{figure}[t]
		\includegraphics[width=\linewidth]{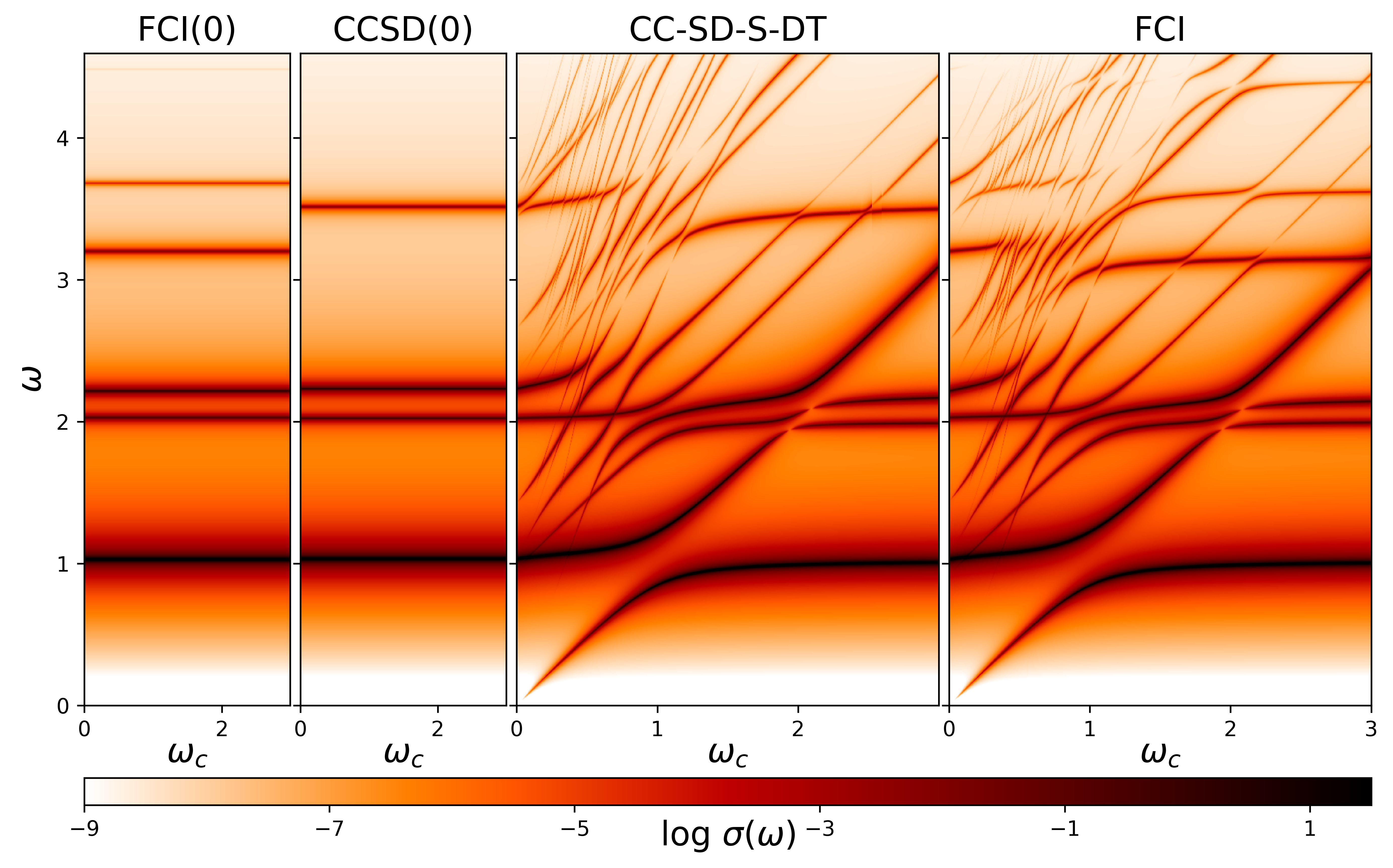}
		\caption{Ground-state absorption cross section $\sigma (\omega)$ of the half-filled four-site Hubbard chain in a cavity as a function of cavity frequency $\omega_c$ in the ultra-strong-coupling regime $\gamma=0.2$
			for different levels of CC theory compared with FCI results (right-hand side) and zero-coupling limit (FCI-0, left-hand side). Parameters: $t_0 = 0.5$, $U= 1.0$, $d = [-1.5, -0.5, 0.5, 1.5]$, $n_\mathrm{max}=7$, $\eta = 0.005$. \label{fig:spectra0-2}}
	\end{figure}
	
	The exact (FCI) spectrum in the ultra-strong coupling case is much more complicated (Fig.~\ref{fig:spectra0-2}).
	Additional features are seen beyond the simple
	combination of non-dispersive matter lines and linear dispersive photon lines with avoided crossings. 
	These include induced transparencies (dark states) in regions of crossings of various absorption lines with the lowest one-photon line, 
	and complicated structures in the high energy part of the spectrum where many multi-photon processes overlap in the spectrum.
	
	The CC-SD-S-DT calculation captures much of this complex structure, with remaining deviations mainly caused by the truncation of the electronic cluster operator.
	The low-energy part of the spectrum is reproduced extremely accurately, including the induced transparencies. The qualitative features of the high-energy part of the spectrum are also captured.
	
	Overall, we have shown that subtle light-matter correlations that appear in strong and ultra-strong cavity experiments can be captured in the framework of CC theory, paving the way for high-accuracy modeling and interpretation of experiments in these regimes. 
	
	Although this work has focused on model systems, extension to real \emph{ab initio} Hamiltonians is straightforward. In this work we simulated the CC equations in the framework of exact diagonalization, but the formulation of the theory in terms of photon excitation operators $\hat\tau_n$ means that affordable polynomial scaling implementations will very closely mirror standard electronic CC codes, but with extra channels to describe the additional quantum degrees of freedom.
	
	The individual photonic creation and annihilation operators are quadratic expressions in the excitation operators $\hat\tau_n$ (for example $\hat a=\sum_n \sqrt n\, (\hat{\tau}_{n-1}\hat{\tau}^\dagger_n)$) so that the light-matter interaction in Eq.~(\ref{eq:Hgen}) becomes a four-point interaction, exactly analogous to the four-point two-particle interaction in the electronic \emph{ab initio} Hamiltonian. Thus the equations for the polynomial-scaling implementation of polaritonic CC-SD-S-DT theory are effectively identical to a subset of the conventional CCSDT equations, but with removal of exchange diagrams and inclusion of alternative values in place of two-electron integrals. The theory scales as $\mathcal{O}(N^6\cdot n_{\rm max})$ that is roughly the same as the $\mathcal{O}(N^6)$ scaling of conventional CCSD theory, since $n_{\rm max}$ is usually much smaller than the number of electronic orbitals $N$.
	
	Furthermore, the formalism developed here can be extended to CC theories for coupling of electrons to polarization modes, phonons or thermal reservoirs, including coupling to multiple boson modes and boson-boson interactions. Work along these lines is underway in our groups.

	\acknowledgments{The authors thank M.~Ruggenthaler and C.~Sch\"afer for helpful discussions.
		The authors are grateful to EPSRC (EP/R014493/1 and EP/R014183/1) and to the European Research Council (ERC-2015-AdG-694097) for funding. UM is supported through IMPRS-UFAST.
		CB is supported through the EPSRC TMCS Centre for Doctoral Training (EP/L015722/1).
		The Flatiron Institute is a division of the Simons Foundation.}


%
%
%

\appendix
\section{Convergence study, photon number cut-off}
\label{app:conv_n}
\renewcommand{\textfraction}{0.1}
\renewcommand{\topfraction}{1}
\renewcommand{\bottomfraction}{1}

\begin{figure}[h!]
	\includegraphics[width=.41\textwidth]{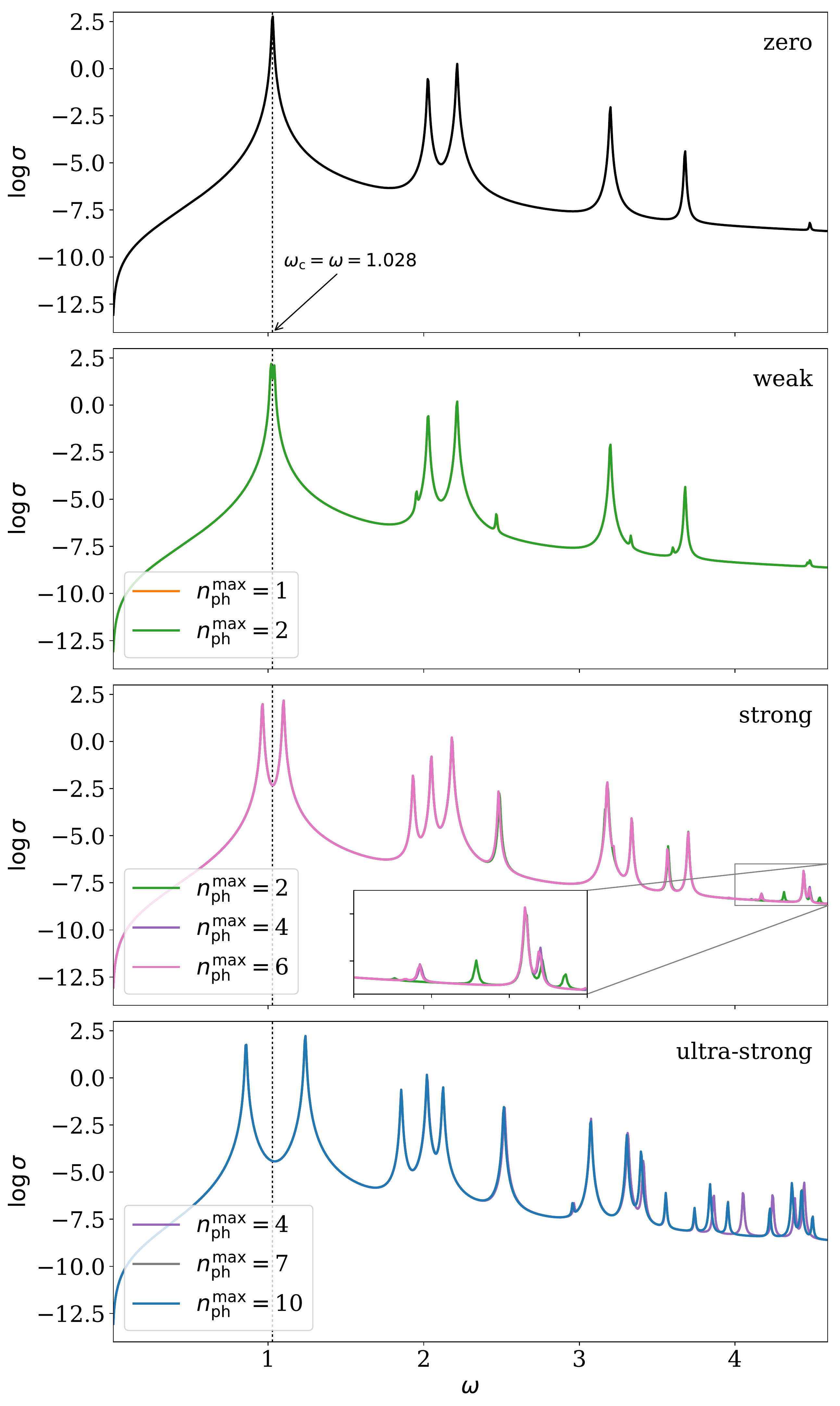}
	\caption{Convergence of the optical spectra with the number of photons for weak ($\gamma=0.01$), strong ($\gamma=0.07$), and ultra-strong ($\gamma=0.2$) coupling compared to the bare-molecule spectrum.
		The following parameters were used: $\omega_c=1.028$, $t_0 = 0.5$, $U= 1.0$, $d = [-1.5, -0.5, 0.5, 1.5]$,  $\eta = 0.005$.}
	\label{fig:phnum}
\end{figure}

\onecolumngrid

\begin{figure*}[h!]
	\includegraphics[width=\textwidth]{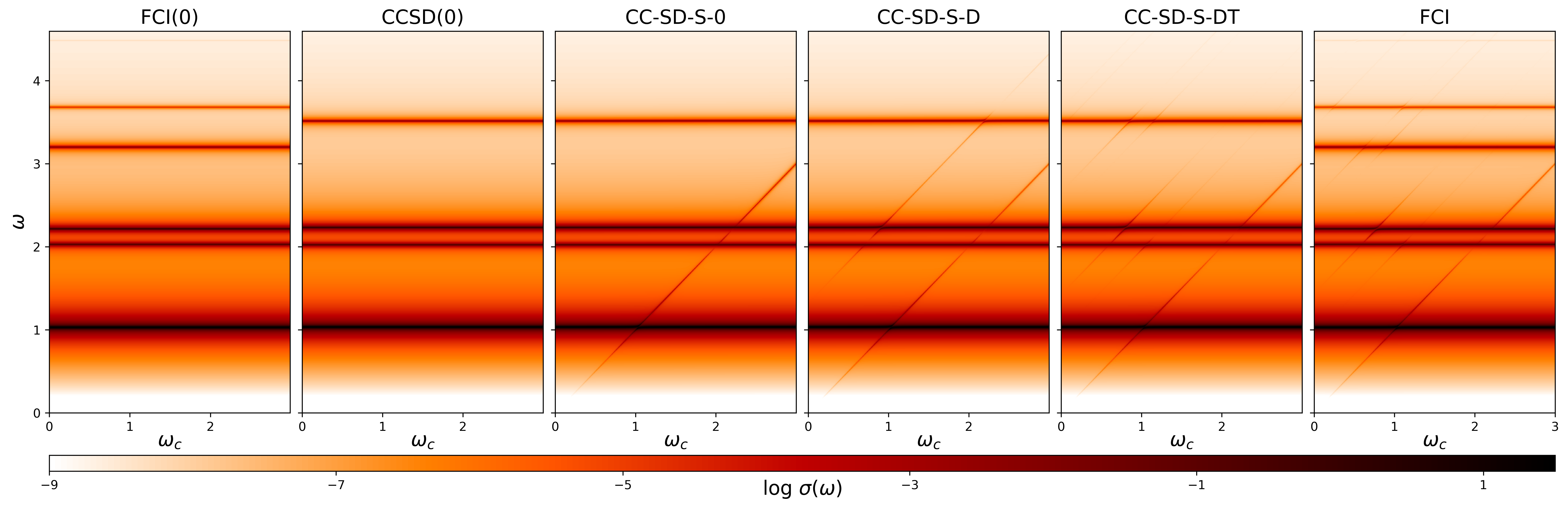}
	\caption{Ground-state absorption cross section $\sigma (\omega)$  of the half-filled four-site Hubbard chain in a cavity as a function of cavity frequency $\omega_c$ in the weak-coupling regime $\gamma=0.01$
		for different levels of CC theory compared with FCI results (right-hand side) and zero-coupling limit (FCI-0, left-hand side). We observe the usual non-dispersive matter absorption lines and additionally some low-intensity liner dispersive branches for one-photon processes. No multi-photon processes or significant Rabi splittings occur, meaning that treating the photon field pertubatively would be sufficient in this regime of light-matter coupling. 
		Parameters: $t_0 = 0.5$, $U= 1.0$, $d = [-1.5, -0.5, 0.5, 1.5]$, $n_\mathrm{max}=1$, $\eta = 0.005$. \label{fig:spectra0-01} }
\end{figure*}

The Fock space for the bosonic modes is infinite, so a reasonable truncation has to be performed in practice by imposing a maximum number of quanta per mode $n^\mathrm{max}_\mathrm{ph}$.
Here, we demonstrate convergence with respect to $n^\mathrm{max}_\mathrm{ph}$ for the FCI spectra of Fig.~\ref{fig:spectra0-05}, Fig.~\ref{fig:spectra0-2}  and Fig.~\ref{fig:spectra0-01}. The absorption cross sections $\sigma(\omega)$ are plotted in Figure~\ref{fig:phnum} for the cavity frequency $\omega_c = 1.028$, where  the cavity is resonant to the first absorption peak of the bare electronic system, and for different photonic cut-offs $n^\mathrm{max}_\mathrm{ph}$ for weak, strong, and ultra-strong coupling and compared to the bare spectrum. We observe a convergence of the spectra at $n^\mathrm{max}_\mathrm{ph}=1$ (weak coupling), $n^\mathrm{max}_\mathrm{ph}=4$ (strong coupling) and $n^\mathrm{max}_\mathrm{ph}=7$ (ultra-strong coupling), respectively. These values of $n^\mathrm{max}_\mathrm{ph}$ were used for all results presented in this paper.

\section{Absorption cross section, weak-coupling regime}
\label{app:weak}

\onecolumngrid
\section{Correction method for close-lying eigenstates}
\label{app:corr}
Coupled-cluster EOM provides a powerful framework for treating excited states that is systematically improvable towards exact solutions to the Schr\"odinger equation. 
One downside is that at finite truncation of the cluster operator, the theory is non-hermitian, and this can lead to difficulties for near-degenerate excitations. 
%
For example, 
the nonhermitian model Hamiltonian ($\hat{\bar H}$) has different right and left eigenvectors, and pairs of right (or left) eigenvectors associated with a near-degeneracy can
become almost parallel. Secondly, the eigenvalues of $\hat{\bar H}$ can develop non-zero imaginary components. 

The issues can be resolved based on the analysis of K\"ohn and Tajti \cite{kohn2007can}, and we refer the interested reader to their work for a detailed discussion.
Here we briefly outline their method to fully specify the calculations we present.

The eigenvalue $E_i$ of the similarity-transformed Hamiltonian $\hat{\bar{H}}$ is associated with left and right eigenvectors $\bra{\widetilde{\Psi}_i}$ and $\ket{\Psi_i}$ that fulfill the bi-orthogonality condition
\begin{equation}
\braket{\widetilde{\Psi}_i}{ \Psi_j} = \delta_{ij}.
\end{equation}
The right eigenvectors are conventionally normalized, but generally nonorthgonal.
The elements of the metric matrix $\mathbf S$ are $S_{ij} = \braket{\Psi_i}{ \Psi_j}$, so we have $S_{ii} = 1$ and $S_{ij}\ne 0$ with $i\ne j$.
The left and right eigenvectors are related via
\begin{equation}
\ket{\widetilde{\Psi}_i} = \sum\limits_j \ket{\Psi_j} \left[ \boldsymbol{S}^{-1}\right]_{ji} \;.
\end{equation}

This relation between the two causes problems, when right vectors become (almost) parallel $S_{ij} \rightarrow 1 $. The norm of the corresponding left eigenvectors diverges and for $S_{ij} = 1 $ the inverse of $\mathbf S$ simply does not exist, as the right vectors $\ket{\Psi_i}$ do not span the full space anymore.

In order to compensate for this behavior, we employ here a correction scheme based on the method of K\"ohn \emph{et al} \cite{kohn2007can}. The idea is basically to rotate the close-lying eigenstates within their subspace such that their overlap becomes smaller and the implications described above become less pronounced.

For simplicity, consider only two close-lying eigenstates of  $\hat{\bar{H}}$
\begin{align}
\hat{\bar{H}} \ket{\Psi_1} & = \varepsilon_1 \ket{\Psi_1}, &
\hat{\bar{H}} \ket{\Psi_2} & = \varepsilon_2 \ket{\Psi_2}. 
\end{align}
We build a subspace matrix and shift it by $\Lambda = \frac{\varepsilon_2 - \varepsilon_1}{2}$
\begin{equation}
\boldsymbol{A} = \left( \begin{matrix}
\avin{\widetilde{\Psi}_1}{\overline{H}}{\Psi_1} & \avin{\widetilde{\Psi}_1}{\overline{H}}{\Psi_2} \\
\avin{\widetilde{\Psi}_2}{\overline{H}}{\Psi_1} & \avin{\widetilde{\Psi}_2}{\overline{H}}{\Psi_2} 
\end{matrix} \right) - \Lambda \mathds{1}
= \left( \begin{matrix}
- \Lambda & 0 \\
0         & \Lambda 
\end{matrix} \right),
\end{equation}
such that $\boldsymbol{A}$ is traceless and has either purely real or purely imaginary eigenvalues.
We further write the metric matrix as
\begin{equation}
\boldsymbol{S} = \left( \begin{matrix}
1 & S \\
S & 1 
\end{matrix} \right)
= \left( \begin{matrix}
1            & \cos \varphi \\
\cos \varphi & 1 
\end{matrix} \right),
\end{equation}
where $S$ is the overlap of the two right vectors $S = \braket{\Psi_1}{\Psi_2} $.
If $S$ is complex, the state $\ket{\Psi_2}$ has to be rotated such that $S$ becomes real.
\begin{align}
\ket{\Psi_2} &\rightarrow e^{i\theta}, & \theta & = \arctan \left( \frac{{\rm Im}\, S}{{\rm Re}\, S}\right).
\end{align}

We now use the overlap matrix $\boldsymbol{S}$ to first orthogonalize the right-hand basis within the subspace spanned by $\ket{\Psi_1}$ and $\ket{\Psi_2}$ by multiplying it with $\boldsymbol{S}^{-1/2}$
\begin{equation}
\boldsymbol{S}^{-1/2} = \frac{1}{2 |\sin \varphi |}\left( \begin{matrix}
\sqrt{1-\cos \varphi }+\sqrt{1 + \cos \varphi} & -\sqrt{1-\cos \varphi }+\sqrt{1 + \cos \varphi} \\
-\sqrt{1-\cos \varphi }+\sqrt{1 + \cos \varphi } & \sqrt{1-\cos \varphi }+\sqrt{1 + \cos \varphi} \\
\end{matrix} \right).
\end{equation}
We then choose a new smaller overlap $\Sigma$ (and the corresponding angle $\vartheta = \arccos \Sigma$). We define it as function of the old overlap $S$ and eigenvalue difference $\Lambda$ and further impose the condition that it is monotonous and is bounded from below by 0 and from above by a maximum value $\Sigma_{\rm max}$. The specific form of the function is not that important and we take the one used in \cite{kohn2007can}.
\begin{equation}
\Sigma = \Sigma_{\rm max} \cdot \begin{cases}
\tanh \left( |S|/\Sigma_{\rm max}\right), & \text{if } \Lambda \text{ real} \\
\tanh \left( 1/ (|S|\cdot \Sigma_{\rm max}) \right), & \text{if } \Lambda \text{ imaginary.} 
\end{cases}
\end{equation}
The orthogonalized eigenstates are then rotated again, this time to have the new overlap $\Sigma = \cos \vartheta$. The corresponding rotation matrix reads
\begin{equation}
\boldsymbol{\Sigma}^{1/2} = \frac{1}{2 |\sin \vartheta |}\left( \begin{matrix}
\sqrt{1-\cos \vartheta }+\sqrt{1 + \cos \vartheta} & \sqrt{1-\cos \vartheta }-\sqrt{1 + \cos \vartheta} \\
\sqrt{1-\cos \vartheta }-\sqrt{1 + \cos \vartheta } & \sqrt{1-\cos \vartheta }+\sqrt{1 + \cos \vartheta} \\
\end{matrix} \right).
\end{equation}

To summarize the derivation above, the new subspace basis is obtained via
\begin{equation}
\left( \begin{matrix} \ket{\Psi_1} \\ \ket{\Psi_2} \end{matrix} \right) \rightarrow  \boldsymbol{\Sigma}^{1/2} \boldsymbol{S}^{-1/2} \left( \begin{matrix} \ket{\Psi_1} \\ \ket{\Psi_2} \end{matrix} \right).
\end{equation}
Then, the vectors $\ket{\Psi_1}, \ket{\Psi_2}$ have to be normalized and the left eigenvectors have to be transformed accordingly.

So far, we have just rotated the eigenstates within their subspace and the eigenvalues remained unaffected by these operations. 
Now, also the eigenvalues have to be adapted with respect to the new overlap. This is done with following formula
\begin{equation}
\Lambda_{\rm corr} = \pm \Lambda \frac{| \sin \vartheta |}{|\sin \varphi |} \cdot \begin{cases}
1, & \text{if } \Lambda \text{ real} \\
-i S, & \text{if } \Lambda \text{ imaginary.}
\end{cases}
\end{equation}
We refer the reader to reference \cite{kohn2007can} for a detailed derivation.
Throughout the paper we use $\Sigma_{\rm max}=0.2$ and apply the correction scheme described above as soon as $\Lambda < 0.05$ or $\Lambda$ imaginary. These parameters can be adapted if needed.

\twocolumngrid
%

\end{document}